\documentclass{moriond}

\def\Journal#1#2#3#4{{#1} {\bf #2}, #3 (#4)}

\def\PRL{\em Phys. Rev. Lett.}
\def\PRD{{\em Phys. Rev.} D}

\def\CQG{\em Class. Quant. Grav.}
\def\GRG{\em Gen. Rel. Grav.}

\def\al{\alpha}
\def\be{\beta}
\def\ga{\gamma}
\def\de{\delta}
\def\ep{\epsilon}

\def\ze{\zeta}
\def\et{\eta}

\def\ka{\kappa}
\def\la{\lambda}

\def\rh{\rho}

\def\ch{\chi}

\def\om{\omega}

\def\cG{{\cal G}}
\def\cl{{\cal L}}

\def\fr#1#2{{{#1} \over {#2}}}
\def\half{{\textstyle{1\over 2}}}

\def\frac#1#2{{\textstyle{{#1}\over {#2}}}}

\def\lsim{\mathrel{\rlap{\lower4pt\hbox{\hskip1pt$\sim$}}
   \raise1pt\hbox{$<$}}}
\def\gsim{\mathrel{\rlap{\lower4pt\hbox{\hskip1pt$\sim$}}
   \raise1pt\hbox{$>$}}}

\def\prt{\partial}

\def\etal{{\it et al.}}
\def\pt#1{\phantom{#1}}

\newcommand{\beq}{\begin{equation}}
\newcommand{\eeq}{\end{equation}}
\newcommand{\bea}{\begin{eqnarray}}
\newcommand{\eea}{\end{eqnarray}}
\newcommand{\bit}{\begin{itemize}}
\newcommand{\eit}{\end{itemize}}
\newcommand{\rf}[1]{(\ref{#1})}

\def\kr{{(k^{(4)})}{}}
\def\kdr{{(k^{(5)})}{}}
\def\kddr{{(k^{(6)}_1)}{}}
\def\krr{{(k^{(6)}_2)}{}}

\def\kb{\overline{k}{}}
\def\kt{\widetilde{k}{}}

\def\kbddr{{(\kb^{(6)}_1)}{}}
\def\kbrr{{(\kb^{(6)}_2)}{}}

\def\tb{\overline{t}}

\def\sb{\overline{s}}

\def\sbddr{{(\sb^{(6)}_1)}{}}
\def\sbh{\widehat{\overline{s}}{}}

\def\ub{\overline{u}}

\def\ubddr{{(\ub^{(6)}_1)}{}}
\def\ubh{\widehat{\overline{u}}}

\def\kl{{\ka\la}}
\def\mn{{\mu\nu}}
\def\ab{{\al\be}}
\def\abl{{\al\be\ldots}}
\def\abgd{{\al\be\ga\de}}
\def\abkl{{\al\be\ka\la}}
\def\gdmn{{\ga\de\mu\nu}}
\def\klmn{{\ka\la\mu\nu}}

\def\cld#1{\cl^{(#1)}_{\rm LV}}
\def\cG{{G}}

\def\mbf#1{{\bf #1}}

\newcommand{\Photo}{\includegraphics[height=35mm]{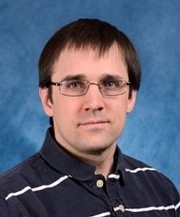}}

\begin{document}
\vspace*{4cm}
\title{What do we know about Lorentz Symmetry?}

\author{ Q.G.\ Bailey }

\address{Department of Physics, 
Embry-Riddle Aeronautical University, 3700 Willow Creek Road,\\
Prescott, AZ 86301, USA}

\maketitle\abstracts{
Precision tests of Lorentz symmetry have
become increasingly of interest to the broader 
gravitational and high-energy physics communities. 
In this talk, 
recent work on violations of local Lorentz invariance in gravity is discussed, 
including recent analysis constraining Lorentz violation in a variety of gravitational tests.
The arena of short-range tests of gravity is highlighted, 
demonstrating that such tests are sensitive to a broad class of 
unexplored signals that depend on sidereal time and the geometry of the experiment.}

\section{Overview}

The Einstein equivalence principle
is a crucial founding principle of General Relativity.
The weak equivalence principle (WEP) and local Lorentz invariance (LLI)
are two essential parts of this principle.
The WEP states that gravity acts in a flavor independent manner,
and local Lorentz invariance states that the local symmetries of nature
include rotations and boosts.
Strong experimental support for the WEP and LLI
is necessary for developing a deep understanding of gravity.

Tests of the WEP are abundant,
while tests of local Lorentz invariance have been 
largely limited to the matter sector.\cite{cw}
Though the latter are primarily confined to the flat space-time limit, 
the breadth and scope of the current experimental tests of Lorentz invariance
is impressive.\cite{tables}
The motivation for the recent boom in Lorentz symmetry tests in the past two decades 
is due not only to the importance of this principle as a foundation of modern physics
but also to the intriguing possibility that minuscule violations
of Lorentz symmetry may occur in nature as a signal of Planck-scale physics.\cite{ksp,jtreview}

When definitive knowledge of the underlying physics is lacking, 
the method of effective field theory is a powerful tool for investigating 
physics at experimentally relevant scales.
For studying local Lorentz invariance in gravity,  
effective field theory is particularly well suited.
Using a Lagrange density containing the usual Einstein-Hilbert term, 
together with a series of observer scalar terms, 
each of which is constructed by contracting coefficient fields with gravitational field operators 
of increasing mass dimension $d$,
one constructs the gravity sector of the effective field theory describing
general local Lorentz violations for spacetime-based gravitation.\cite{akgrav}
One can also consider a series of terms involving matter-gravity couplings
where Lorentz-violating terms from the flat space-time scenario are coupled to gravity,
thereby imparting observability to some lagrangian terms that are unobservable in flat space-time.\cite{kt}
To date, 
the so-called minimal sector of this framework,
consisting of terms with operators of the lowest mass dimension $d \leq 4$,
has been explored in experimental searches for local Lorentz violation 
and phenomenological studies in gravity related tests.\cite{2007Battat,MullerInterf,2012Iorio,Hohensee,2013Hees,2013Bailey,2014Shao,2015Long,2015ShaoEtal,bk06,b0910,tb11,bkx15}

It is well known that Newtonian gravity and relativistic corrections from
General Relativity accurately describe the dominant physics
at the typical stellar system level.
Experimental and observational searches for Lorentz violation within the general 
effective field framework described above have focused on observables at this level.
However, 
it is currently unknown whether gravity obeys Newton's law of gravitation
on small scales below about 10 microns.
In fact, 
it is within the realm of possibility that forces vastly stronger than 
the usual Newtonian inverse-square law could exist.
In a recent work,
a systematic study of local Lorentz violation with $d>4$ has been initiated.\cite{bkx15}
Lorentz-violating corrections to the Newtonian force law vary as $1/r^{d-2}$,
since lagrangian terms constructed with operators 
of higher mass dimension $d$ involve more derivatives.
The sharpest sensitivities to effects from operators with $d>4$ are therefore most likely 
to come from short-range tests of gravity. 
The phenomenology of such signals are discussed in this presentation.

\section{Gravity Sector}

It is known that explicit Lorentz violation 
is generically incompatible with Riemann geometry
or is technically unnatural in spacetime theories of gravity,
so we focus here on spontaneous violation of Lorentz symmetry.\cite{ksp,akgrav,bl15}
Spontaneous Lorentz violation occurs 
when an underlying local Lorentz invariant action involves 
gravitational couplings to tensor fields $k_\abl$
that acquire nonzero background values $\kb_\abl$.\cite{fn1}
The resulting phenomenology violates local Lorentz invariance
due to the presence of nonzero backgrounds and so the backgrounds $\kb_\abl$ 
are called coefficients for Lorentz violation.\cite{sme}
The massless Nambu-Goldstone and massive modes associated with
spontaneous breaking are contained in the field fluctuations $\kt_\abl\equiv k_\abl - \kb_\abl$
and can potentially impact the physics.

The Lagrange density of the effective field theory action, 
focusing on pure gravitational and matter gravity couplings,
can be written as the sum of four terms,
\beq
\cl = \cl_{\rm EH} + \cl_{\rm LV} + \cl_k + \cl_{\rm M}.
\eeq
The first term is the usual Einstein-Hilbert term is $\cl_{\rm EH}= \sqrt{-g}R /16\pi G_N$,
where $G_N$ is Newton's gravitational constant,
while the second term $\cl_{\rm LV}$ contains the Lorentz-violating couplings. 
The dynamics of the coefficient fields triggering the spontaneous Lorentz violation
are contained in $\cl_k$.
Finally, 
the matter is described by $\cl_{\rm M}$.

We can also include into the matter sector, 
the so-called matter-gravity couplings.\cite{kt}
These terms are determined from a general Lorentz-violating lagrangian series for Dirac fermions.
For classical tests in which spin is irrelevant the physical effects can be shown
to be equivalent to a classical action for point particles of the form
\beq
S_{\rm M, LV} = -\int d\la \left( m \sqrt{-(g_\mn +2c_\mn) u^\mu u^\nu} -a_\mu u^\mu \right),
\label{mg}
\eeq
where $u^\mu=dx^\mu/d\la$ is the four velocity of the particle and $c_\mn$ and $a_\mu$
are the species-dependent coefficients for Lorentz violation
that also effectively violate the WEP.
Observables for Lorentz violation from this action involve
a variety of signals in terrestrial and space-based gravitational tests as well as
solar system observations and beyond.\cite{jtreview}
In particular, 
experiments designed to test the WEP are ideally suited
to measure the coefficients $a_\mu$ and $c_\mn$.\cite{wep}
 
In the pure-gravity sector, 
a series involving observer covariant gravitational operators comprise the term $\cl_{\rm LV}$:
\beq
\cl_{\rm LV} = \fr {\sqrt{-g}}{16\pi G_N}
(\cld4 +\cld5 +\cld6 + \ldots ),
\eeq
Each subsequent term involves higher mass dimension $d$ and is
formed by contracting covariant derivatives $D_\al$ and curvature tensors $R_{\abgd}$
with the coefficient fields $k_\abl$.
Though much of the discussion can be generalized to $d>6$,
here,
we consider terms with $4\leq d\leq 6$.

The first term in the series with $d=4$ is known as
the minimal term $\cld4$ given by 
\beq
\cld4 = \kr_{\abgd} R^{\abgd},
\label{minimallag}
\eeq
where the coefficient field $\kr_{\abgd}$ is dimensionless.\cite{akgrav}
Due to the contraction with to the Riemann tensor, 
$\kr_{\abgd}$ has the index symmetries of the Riemann tensor.
In particular, the $20$ independent coefficients can be decomposed 
into a traceless part $t_{\abgd}$ with $10$ coefficients,
a trace $s_{\ab}$ with $9$ coefficients,
and the double trace $u$.

In the linearized limit of gravity, 
assuming an origin in spontaneous symmetry breaking,
the vacuum value of the coefficient $\ub$ acts as an 
unobservable rescaling of Newton's constant $G_N$.
In contrast, 
many phenomenological effects are generated by the $\sb_{\ab}$ coefficients.\cite{bk06,b0910,tb11,2013Bailey}
These coefficients have been constrained to various degrees to parts in $10^{10}$
by numerous analyses using including
lunar laser ranging,
atom interferometry,
short-range tests,
satellite ranging,
light bending and orbital simulations,
precession of orbiting gyroscopes,
pulsar timing and spin precession,
and solar system ephemeris.\cite{2007Battat,MullerInterf,2015Long,2015ShaoEtal,2012Iorio,2013Hees,2013Bailey,2014Shao,2012Iorio}
At leading order in the linearized gravity limit, 
the coefficients in $\tb_{\abgd}$ are absent.
The physical effects of these 10 independent coefficients remain unknown.\cite{2015bonder}

For the mass dimension $5$ term, 
using covariant derivatives and curvature the general expression is  
\beq
\cld5 = \kdr_{\abgd\ka} D^\ka R^{\abgd}.
\eeq 
The coefficient fields $\kdr_{\abgd\ka}$ can be shown to contain
$60$ independent quantities by using the properties of the coupling 
with the covariant derivative and the Riemann tensor.
Some features of this term can be determined from its space-time symmetries.
Under the operational definition of the CPT transformation,
the expression $D^\ka R^{\abgd}$ is CPT odd.\cite{akgrav}
This can have profound effects for phenomenology.
For example, 
in the nonrelativistic limit the associated Newtonian gravitational force from $\cld5$
would receive pseudovector contributions rather than conventional vector ones.
Self accelerations of localized bodies would then occur due to these coefficients.
In other sectors, 
some CPT-odd coefficients with similar issues are known.\cite{km09}
For the higher mass dimension terms, 
the initial focus is on (stable) corrections to the Newtonian force and
so the phenomenology of these coefficients, 
at higher post-newtonian order, 
remains an open issue.

For the mass dimension six terms, 
the coefficient fields are contracted with appropriate powers
of curvatures and covariant derivatives,
thus we write $\cld6$ in the form
\beq
\cld6 = \half \kddr_{\abgd\kl} \{D^\ka, D^\la\}  R^{\abgd} + \krr_{\abgd\klmn} R^{\klmn} R^{\abgd}.
\label{lvlag}
\eeq
In natural units, 
the coefficient fields $\kddr_{\abgd\kl}$ and $\krr_{\abgd\klmn}$
have dimensions of squared length,
or squared inverse mass.
Since the commutator of covariant derivatives is directly related to curvature, 
the anticommutator of covariant derivatives suffices for generality in the first term.
The first and last four indices on $\krr_{\abgd\klmn}$
inherit the symmetries of the Riemann tensor
as do the first four indices on $\kddr_{\abgd\kl}$.
A cyclic-sum condition of the form $\sum_{(\ga\de\ka)} \kddr_{\abgd\kl} = 0$
applies due to the Bianchi identities.  
These tensor symmetry conditions can be used to determine that 
there are 126 and 210 independent components in 
$\kddr_{\abgd\kl}$ and $\krr_{\abgd\klmn}$, respectively.

In an underlying theory,
Lorentz-violating derivative couplings of fields to gravity 
could give rise to the coefficients $\kddr_{\abgd\kl}$. 
It is straightforward to construct models that produce this type of coupling,
although examples are currently unknown to us in the literature.
On the other hand, 
in many models specific forms of quadratic Lorentz-violating couplings 
occur as a result of integrating over fields in the underlying action 
that have Lorentz-violating couplings to gravity.
General quadratic Lorentz-violating curvature couplings
are represented by the coefficients $\krr_{\abgd\klmn}$, 
thus including various models as special cases.
For example, 
models of this type include include the cardinal model,
various types of bumblebee models, 
and Chern-Simons gravity.\cite{jp,ay,cardinal,bumblebee,akgrav}
It is also useful to note the implications of introducing these higher derivative terms.  
It is well known that lagrangian terms with higher than
two derivatives can suffer from stability issues.
However, 
in the effective field theory formalism here, 
these terms with higher derivatives are to be considered
only in the perturbative limit, 
thus they are assumed small compared to the conventional
terms with only two derivatives.

To extract the linearized modified Einstein equation
resulting from the terms \rf{lvlag},
we assume an asymptotically flat background metric $\et_\ab$ as usual, 
and write the background coefficients as
$\kbddr_{\abgd\kl}$ and $\kbrr_{\abgd\klmn}$.
The analysis is performed at linear order in the 
metric fluctuation $h_\ab$ 
and we seek results to leading order in the coefficients
(assuming they are small).
The coefficients are are assumed constant in asymptotically flat coordinates.
We can re-express the contributions of the fluctuations $\kt_\abl$ 
in terms of the metric fluctuations and the background coefficients
by imposing the underlying diffeomorphism invariance on the dynamics
and that the conservations laws must hold (i.e., covariant
conservation and symmetry of the energy-momentum tensor).
This procedure yields a modified Einstein equation
expressed in terms of $\kb_\abl$ and quantities involving $h_\ab$ such 
as the linearized curvature tensor.
Similar procedures are detailed in the literature.\cite{bk06,kt}
To establish signals for local Lorentz violation in specific experiments,
the phenomenology of the modified equation can be studied.
An interesting feature of the coefficient fields 
$\krr_{\abgd\klmn}$ is that for the linearization outlined above 
the coefficient fluctuations can be neglected
because these contribute only at nonlinear order.
This feature did not occur in the minimal, 
mass dimension $4$ case.\cite{bk06}

Following the procedure above the linearized modified Einstein equations can be obtained,
after some calculation,
and they can be written in the compact form
\bea
G_\mn &=& 8\pi G_N (T_M)_\mn 
-2\sbh^{\al\be} \cG_{\al(\mu\nu)\be}
-\frac 12 \ubh G_\mn
+a \kbddr_{\al(\mu\nu)\be\ga\de} \prt^\al \prt^\be R^{\ga\de}
\nonumber\\
&&
+ 4 \kbrr_{\al\mu\nu\be\ga\de\ep\ze} \prt^\al \prt^\be R^{\ga\de\ep\ze},
\label{lineq3}
\eea
where double dual of the Riemann tensor is
$\cG_\abgd \equiv \ep_\abkl \ep_\gdmn R^\klmn/4$ and the Einstein tensor is
$G_{\al\be} \equiv \cG^{\ga}_{\pt{\ga}\al\ga\be}$.
All gravitational tensors are understood to be linearized in $h_\mn$
in Eq.\ \rf{lineq3}.
For notational convenience, 
the ``hat" notation is used for the following operators:
\bea 
\ubh &=& -2\ub + \ubddr_{\al\be} \prt^\al \prt^\be,
\nonumber\\ 
\sbh_{\al\be} &=& \frac 12 \sb_{\al\be} + \sbddr_{\abgd} \prt^\ga \prt^\de,
\label{hats}
\eea
where
$\ubddr_{\ga\de} \equiv \kbddr^{\al\be}_{\pt{\al\be}\abgd}$
and 
$\sbddr^{\al}_{\pt{\al}\be\ga\de} \equiv 
\kbddr^{\al\ep}_{\pt{\al\ep}\be\ep\ga\de} 
- \de^\al_{\pt{\be}\be} \ubddr_{\ga\de}/4 $.
The factors in front of the $\ub$ and $\sb$ are chosen to match earlier work
in the mass dimension $4$ case.
For the $d=4$ Lorentz-violating term \rf{minimallag},
the entire contribution is contained in $\ubh$ and $\sbh_{\al\be}$.
There are also $d=6$ terms contained in $\ubh$ and $\sbh_{\al\be}$.
A model-dependent real number $a$ remains in Eq.\ \rf{lineq3}
that depends on the underlying dynamics specified by the 
Lagrange density $\cl_k$.
Furthermore, 
the quantity $a$ may be measurable 
independently of the coefficients $\kbddr_{\abgd\kl}$ 
and $\kbrr_{\abgd\klmn}$, 
revealing a way to extract information about the dynamics
behind spontaneous Lorentz symmetry breaking, 
should it occur in nature.

Numerous phenomenological consequences
both for relativistic effects, including gravitational waves,
and effects in post-newtonian gravity are likely to be implied 
by the modified Einstein equation \rf{lineq3}.
Since we expect the mass dimension $6$ terms
to be dominant on short distance scales, 
we consider the nonrelativistic limit and assume a 
source with mass density $\rh(\mbf r)$.
In this limit, 
a modified Poisson equation is revealed:
\beq
-\vec \nabla^2 U = 
4\pi G_N \rh 
+(\kb_{\rm eff})_{jk} \prt_j \prt_k U
+(\kb_{\rm eff})_{jklm} \prt_j \prt_k \prt_l \prt_m U,
\label{newt}
\eeq
where the modified Newton gravitational potential is $U(\mbf r)$.
The effective coefficients for Lorentz violation 
with totally symmetric indices in this equation are 
$(\kb_{\rm eff})_{jk}$ and $(\kb_{\rm eff})_{jklm}$.
The former are associated with mass dimension $4$ and are related to the $\sb_{00}$,
$\sb_{jk}$ and $\ub$ coefficients and are detailed in Ref.\ 16, 
while the latter depend on the mass dimension $6$ coefficients
and are the primary focus of more recent work.
The effective coefficients $(\kb_{\rm eff})_{jklm}$ are linear combinations 
of the $d=6$ coefficients $\kddr_{\abgd\kl}$ and $\krr_{\abgd\klmn}$.
Since it is largely irrelevant for present purposes, 
we omit the explicit lengthy form of this relationship.
Nonetheless it is important to note that many of the independent components
$\kddr_{\abgd\kl}$ and $\krr_{\abgd\klmn}$ appear.

With the Lorentz-violating term assumed to generate
a small correction to the usual Newtonian potential, 
we can adopt a perturbative approach
to solve the modified Poisson equation \rf{newt}.
On the length scales of experimental interest,
the $d=6$ Lorentz-violating term \rf{lvlag} represents a perturbative correction 
to the Einstein-Hilbert action, 
thus the perturbative approach is consistent with this method of solution. 
Though it involves theoretical complexities that lie outside the present scope, 
the nonperturbative scenario with $\cld6$ dominating the physics could in principle also be of interest.

The solution to the modified Poisson equation \rf{newt} for $d=6$,  
within the perturbative assumption, 
is given by
\beq
U (\mbf r ) =  
G_N \int d^3 r^\prime 
\fr{\rh (\mbf r^\prime )}{ |\mbf r - \mbf r^\prime|}
\left( 1 + \fr { \kb (\widehat{\mbf R}) } {
|\mbf r - \mbf r^\prime|^2} \right) 
+{\frac 45 \pi G_N} \rh( \mbf r ) (\kb_{\rm eff})_{jkjk}.
\label{ULV2}
\eeq
In addition to the conventional Newtonian potential, 
\rf{ULV2} contains a Lorentz-violating 
correction term that varies with the inverse cube of the distance.
Adopting the convenient notation
for the unit vector $\widehat {\mbf R} = (\mbf r - \mbf r^\prime)/|\mbf r - \mbf r^\prime|$,
the anisotropic combination of coefficients $\kb = \kb (\hat{\mbf r})$
is a function of $\hat{\mbf r}$ given by
\beq
\kb (\widehat{\mbf r}) = \frac 32  (\kb_{\rm eff})_{jkjk}
- 9  (\kb_{\rm eff})_{jkll} \hat{r}^j \hat{r}^k 
+ \frac {15}{2} (\kb_{\rm eff})_{jklm} \hat{r}^j \hat{r}^k \hat{r}^l \hat{r}^m.
\label{tildekb}
\eeq
In parallel with the usual dipole contact term in electrodynamics,
the final piece in \rf{ULV2} is a contact term that becomes a delta function 
in the point-particle limit.
Interestingly this last term is absent for the
mass dimension $4$ solution, 
showing up only starting at mass dimension $6$.
Via the Newtonian gravitational field $\mbf g = \mbf\nabla U$,
an inverse-quartic gravitational field results
from the inverse-cube behavior of the potential.\footnote{For this analysis, 
we assume a conventional matter sector with the acceleration of test bodies being $\mbf a = \mbf g$.  
This can be generalized to include effects from other sectors.\cite{kt}} 
Short-range gravity tests measure the deviation from the Newton gravitational force
between two masses, 
and the rapid growth of the force at small distances
suggests that the best sensitivities to Lorentz violation could be achieved in experiments of
this type.\cite{review}

\section{Short-range gravity tests}

Sensitivity to the coefficients $(\kb_{\rm eff})_{jklm}$
occurs instantaneously through the measurements of the force between two masses 
in an Earth-based laboratory frame.
The Earth's rotation about its axis and revolution about the Sun
induce variations of these coefficients with sidereal time $T$, 
since the laboratory frame is noninertial.
The Sun-centered frame is the canonical frame adopted for reporting results
from experimental searches for Lorentz violation.\cite{tables,sunframe}
In this frame, 
$Z$ points along the direction of the Earth's rotation
and the $X$ axis points towards the vernal equinox 2000.
To relate the laboratory frame $(x,y,z)$ to the Sun-centered frame $(X,Y,Z)$, 
a time-dependent rotation $R^{jJ}$ is used if we neglect the Earth's boost 
(which is of order $10^{-4}$), 
where $j = x,y,z$ and $J=X,Y,Z$.
In terms of constant coefficients $(\kb_{\rm eff})_{JKLM}$ in the Sun-centered frame,
the $T$-dependent coefficients $(\kb_{\rm eff})_{jklm}$ in the laboratory frame are given by
\beq
(\kb_{\rm eff})_{jklm} = R^{jJ} R^{kK} R^{lL} R^{mM} (\kb_{\rm eff})_{JKLM}.
\label{rot}
\eeq

One standard commonly adopted is
to take the laboratory $x$ axis pointing to local south,
the $z$ axis pointing to the local zenith.
This convention yields the following rotation matrix:
\beq
R^{jJ}=\left(
\begin{array}{ccc}
\cos\ch\cos\om_\oplus T
&
\cos\ch\sin\om_\oplus T
&
-\sin\ch
\\
-\sin\om_\oplus T
&
\cos\om_\oplus T
&
0
\\
\sin\ch\cos\om_\oplus T
&
\sin\ch\sin\om_\oplus T
&
\cos\ch
\end{array}
\right).
\label{rotmat}
\eeq
The Earth's sidereal rotation frequency is 
$\om_\oplus\simeq 2\pi/(23{\rm ~h} ~56{\rm ~min})$
and the angle $\ch$ is the colatitude of the laboratory.
The modified potential $U$ and the force between two masses 
measured in the laboratory frame will vary with time $T$
as a result of the sidereal variation of the laboratory-frame coefficients.

One simple application
is the point-mass $M$ modified potential. 
To extract the time dependence, 
Eq.\ \rf{rot} is used to express the combination ${\kb (\hat {\mbf r},T)}$
in Eq.\ \rf{tildekb} in terms of coefficients $(\kb_{\rm eff})_{JKLM}$ 
in the Sun-centered frame.
For points away from the origin, 
the potential then takes the form
\beq
U (\mbf r,T) =  \fr {G_N M}{r} 
\left( 1 + \fr {\kb (\hat {\mbf r},T)}{r^2} \right).
\label{ULV3}
\eeq
This contains novel signals in short-range experiments,
where the modified force depends both on direction and sidereal time.
In particular, 
the effective gravitational force between two bodies 
can be expected to vary with frequencies up to and including 
the fourth harmonic of $\om_\oplus$ due to the time dependence in Eq.\ \rf{rot}. 

An asymmetric dependence of the signal on the shape of the bodies
is implied by the direction dependence 
of the laboratory-frame coefficients $(\kb_{\rm eff})_{jklm}$.
In conventional Newton gravity,
the force on a test mass at any point above an infinite plane
of uniform mass density is constant,
and this result remains true for the potential \rf{ULV3}.
However,
it is typically necessary to determine 
the potential and force via numerical integration
for the finite bodies used in experiments.
It turns out that shape and edge effects
play an critical role in determining the sensitivity 
of the experiment to the coefficients for Lorentz violation, 
as suggested by some simple simulations for experimental configurations 
such as two finite planes or a plane and a sphere.\cite{colorado03,iupui03,wuhan}

An anisotropic inverse-cube correction to the usual Newtonian result is involved
in the modified potential \rf{ULV3}.
Existing experimental limits on spherically symmetric inverse-cube potentials 
cannot be immediately converted into constraints on the coefficients $(\kb_{\rm eff})_{JKLM}$.
This is due to the time and orientation dependence of the Lorentz-violating signal,
whereas typical experiments collect data over an extended period 
and disregard the possibility of orientation-dependent effects. 
Thus new experimental analyses will be required
for establishing definitive constraints on the coefficients
$(\kb_{\rm eff})_{JKLM}$ for Lorentz violation.

It is useful to identify a measure of the reach of a given experiment,
given the novel features of short-range tests of local Lorentz violation in gravity
and the wide variety of experiments in the literature.
Generally, 
a careful simulation of the experiment is required,
but rough estimates can be obtained by comparing
the Lorentz-violating potential with the potential modified by a 
two parameter ($\al$,$\la$) Yukawa-like term, 
$U_{\rm Yukawa} = {G_N M} (1+ \al e^{-r/\la})/r$,
which is commonly used for experiments testing short-range gravity.
Sensitivities to Lorentz violation of order $|{\kb (\hat {\mbf r},T)}| \approx \al \la^2/e$
are indicated by comparing the Yukawa form with the potential \rf{ULV3} 
assuming distances $r\approx \la$.
Thus using Eq.\ \rf{tildekb}, 
the sensitivity to combinations of coefficients is approximately 
\beq
|(\kb_{\rm eff})_{JKLM}| \approx \al \la^2/10.
\label{limit}
\eeq
Note that the experiment must be able to detect the usual Newtonian gravitational force
in order to have sensitivity to the perturbative Lorentz violation considered here. 
This is the case for a subset of experiments reported in the literature.
Also, 
distinct linear combinations of $(\kb_{\rm eff})_{JKLM}$
will be accessed by different experiments.

Experiments at small $\la$ that are sensitive to the usual Newtonian force
are the most interesting short-range experiments within this perspective.
For example,
eq.\ \rf{limit} gives the estimate 
$\al \simeq 10^{-3}$ at $\la \simeq 10^{-3}$ m
for the Wuhan experiment which implies the sensitivity
$|(\kb_{\rm eff})_{JKLM}| \simeq 10^{-10}$ m$^{2}$.\cite{wuhan}
However, 
due to the geometry of this experiment, 
edge effects reduce the sensitivity by about a factor
of $100$ and the limits recently obtained are at the
$10^{-8}$ m$^{2}$ level.
The E\"otWash torsion pendulum experiment, 
which has been used to place limits on isotropic
power law deviations from the inverse square law, 
achieves sensitivity of order $\al \simeq 10^{-2}$ at $\la \simeq 10^{-4}$ m.\cite{eotwash,inversesquare}
Thus suggests Lorentz violation can be measured at the level of
$|(\kb_{\rm eff})_{JKLM}| \simeq 10^{-11}$ m$^{2}$,
in agreement with the estimate from a simple simulation.\cite{bkx15}
Other experiments of interest include
the Irvine experiment which achieved $\al \simeq 3\times 10^{-3}$ at $\la \simeq 10^{-2}$ m,
and should be able to obtain 
$|(\kb_{\rm eff})_{JKLM}| \simeq 3 \times 10^{-8}$ m$^{2}$.\cite{irvine85}
Sitting on the cusp of the perturbative limit,
the Indiana experiment achieves
$\al \simeq 1$ at $\la \simeq 10^{-4}$ m.
Naively, 
we would expect an estimated sensitivity of order 
$|(\kb_{\rm eff})_{JKLM}| \simeq 10^{-9}$ m$^{2}$.\cite{colorado03}
However,
since this test uses flat plates, 
edge effects end up suppressing the sensitivity to the $10^{-7}$m$^2$ level.\cite{2015Long}
There are also many other experiments that can potentially probe for 
the $(\kb_{\rm eff})_{JKLM}$ coefficients, 
including ones discussed at this conference.\cite{srmoriond}

Note that the predicted effects can be quite large while having escaped detection to date
in some gravity theories with violations of Lorentz invariance.\cite{kt}
Because the Planck length $\simeq 10^{-35}$ m lies far below
the length scale accessible to existing laboratory experiments on gravity,
the above estimates suggest terms in the pure-gravity sector with $d>4$
are interesting candidates for these ``countershaded" effects.
In any case,
the Einstein equivalence principle for the gravity sector can be 
established on a firm and complete experimental footing with the types of analysis 
described here.
In particular, 
short-range tests of gravity offer an excellent opportunity
to search for local Lorentz violation involving operators of higher mass dimension.

\section*{Acknowledgements}

Travel and housing to present this work was supported 
by the National Science Foundation under grant number PHY-1402890
and by the Moriond Conference Organizing Committee.

\end{document}